\documentclass[twocolumn,american,showpacs,showkeys,amsmath,amssymb,superscriptaddress,a4, citeautoscript,nobibnotes, aps, prb]{revtex4-1}
\pdfoutput=1
\usepackage{array}
\usepackage{float}
\usepackage{multirow}
\usepackage{amsmath}
\usepackage{amssymb}
\usepackage{babel}
\usepackage{graphicx}
\usepackage{subscript}

\usepackage{color,soul}
\usepackage[unicode=true]{hyperref}
\usepackage{breakurl}
\usepackage{color}

\makeatletter
\renewcommand{\p@subsection}{}
\renewcommand{\p@subsubsection}{}
\makeatother
\makeatletter

\usepackage{multirow}\usepackage{bm}\usepackage{gensymb}\usepackage{threeparttable}

\newenvironment{talign*}
 {\csname align*\endcsname}
 {\endalign}
\newenvironment{talign}
{\align}
{\endalign}

\makeatother
\usepackage{babel}
\begin{document}
\selectlanguage{american}%
\title{Inference of COVID-19 epidemiological distributions from Brazilian hospital data}

\author{Iwona Hawryluk}
\affiliation{MRC Centre for Global Infectious Disease Analysis, Department of Infectious Disease Epidemiology, Imperial College London, UK}
\author{Thomas A. Mellan}
\email{t.mellan@imperial.ac.uk}
\affiliation{MRC Centre for Global Infectious Disease Analysis, Department of Infectious Disease Epidemiology, Imperial College London, UK}
\author{Henrique H Hoeltgebaum}
\affiliation{Department of Mathematics, Imperial College London, UK}
\author{Swapnil Mishra}
\affiliation{MRC Centre for Global Infectious Disease Analysis, Department of Infectious Disease Epidemiology, Imperial College London, UK}
\author{Ricardo P Schnekenberg}
\affiliation{Nuffield Department of Clinical Neurosciences, University of Oxford, UK}
\author{Charles Whittaker}
\affiliation{MRC Centre for Global Infectious Disease Analysis, Department of Infectious Disease Epidemiology, Imperial College London, UK}
\author{Harrison Zhu}
\affiliation{Department of Mathematics, Imperial College London, UK}
\author{Axel Gandy}
\affiliation{Department of Mathematics, Imperial College London, UK}
\author{Christl A. Donnelly}
\affiliation{MRC Centre for Global Infectious Disease Analysis, Department of Infectious Disease Epidemiology, Imperial College London, UK}
\affiliation{Department of Statistics, University of Oxford, UK}
\author{Seth Flaxman}
\email{s.flaxman@imperial.ac.uk}
\affiliation{Department of Mathematics, Imperial College London, UK}
\author{Samir Bhatt}
\affiliation{MRC Centre for Global Infectious Disease Analysis, Department of Infectious Disease Epidemiology, Imperial College London, UK}

\date{\today}

\selectlanguage{american}%
\begin{abstract}
Knowing COVID-19 epidemiological distributions, such as the time from patient admission to death, is directly relevant to effective primary and secondary care planning, and moreover, the mathematical modelling of the pandemic generally. We determine epidemiological distributions for patients hospitalised with COVID-19 using a large dataset ($N=21{,}000-157{,}000$) from the Brazilian Sistema  de  Informação  de  Vigilância  Epidemiológica  da  Gripe database. A joint Bayesian subnational model with partial pooling is used to simultaneously describe the 26 states and one federal district of Brazil, and shows significant variation in the mean of the symptom-onset-to-death time, with ranges between 11.2-17.8 days across the different states, and a mean of 15.2 days for Brazil. We find strong evidence in favour of specific probability density function choices: for example, the gamma distribution gives the best fit for onset-to-death and the generalised log-normal for onset-to-hospital-admission. Our results show that epidemiological distributions have considerable geographical variation, and provide the first estimates of these distributions in a low and middle-income setting. At the subnational level, variation in COVID-19 outcome timings are found to be correlated with poverty, deprivation and segregation levels, and weaker correlation is observed for mean age, wealth and urbanicity. 

\end{abstract}


\maketitle

\section{Introduction \label{sec:Introduction}}
Surveillance of COVID-19 has progressed from initial reports on 31st-Dec-2019 of pneumonia with unknown etiology in Wuhan, China,\cite{who_sitrep_1} to the confirmation of $9,826$ cases of SARS-CoV-2 across $20$ countries one month later,\cite{who_sitrep_11} to the current pandemic of greater than $12$ million confirmed cases and $500,000$ deaths globally to date.\cite{who_sitrep_175}
Early estimates of epidemiological distributions provided critical input that enabled modelling to identify the severity and infectiousness of the disease. The onset-to-death distribution,\cite{donnelly_sars, garske_flu} characterising the range of times observed between the onset of first symptoms in a patient and their death, has for example proved crucial in early estimates of the Infection Fatality Ratio (IFR) \cite{verity_estimates_2020}, and was similarly integral to recent approaches to modelling the transmission dynamics of SARS-CoV-2.\cite{flaxman2020nature, china_first_wave,  dana2020brazilian, severity_wuhan,jombart2020inferring,linton2020incubation}  

Initial estimates of COVID-19 epidemiological distributions necessarily relied on relatively few data points, with the events comprising these distributions occurring a period of time that was short compared to the temporal pathologies of the disease progression, resulting in wide confidence or credible intervals and a sensitivity to time-series censoring effects.\cite{verity_estimates_2020} Global surveillance of the disease over the past $197$ days has provided more data to re-evaluate the time-delay distributions of the disease. In particular, public availability of a large number of patient-level hospital records -- currently over $390,000$ in total -- from the SIVEP-Gripe (\emph{Sistema  de  Informação  de  Vigilância  Epidemiológica  da  Gripe}) database published by Brazil's Ministry of Health (MoH),\cite{SIVEP}  provides an opportunity to make robust statistical estimates of the onset-to-death and other time-delay distributions such as onset-to-diagnosis, length of ICU stay, onset-to-hospital-admission, onset-to-hospital-discharge, onset-to-ICU-admission, and hospital-admission-to-death. 
In this work we fit and present an analysis of these epidemiological distributions, with the paper set out as follows. Section \ref{sec:Method} describes the data used from the SIVEP-Gripe database,\cite{SIVEP} and the methodological approach applied to fit distributions using a hierarchical Bayesian model with partial pooling. Section \ref{sec:Results} provides a description of the results from this study from fitting epidemiological distributions at national and subnational level to a range of probability density functions (PDFs). The results are discussed in Section \ref{sec:Discussion}, including associations with socioeconomic factors, such as education, segregation, and  poverty, and conclusions are given in Section \ref{sec:Conclusions}. 

\section{Methods \label{sec:Method}}
\subsection{Data}

The SIVEP-Gripe database provides detailed patient-level records for all individuals hospitalised with severe acute respiratory syndrome, including all suspected or confirmed cases of severe COVID-19.\cite{SIVEP} 
The records include the date of admission, date of onset of symptoms, state where the patient lives, state where they are being treated, and date of outcome (death or discharge), among other diagnosis related variables. We extracted the data for confirmed COVID-19 records starting on 25th February and considered records in our analysis ending on 7th July. The dataset was filtered to obtain rows for onset-to-death, hospital-admission-to-death, length of ICU stay, onset-to-hospital-admission, onset-to-hospital-discharge, onset-to-ICU-admission and onset-to-diagnosis. Onset-to-diagnosis data were split into the diagnosis confirmed by PCR and those confirmed by other methods, such as rapid antibody and antigen tests, called non-PCR throughout this manuscript. 
Entries resulting in distribution times greater than 133 days were considered a typing error and removed, as the first recorded COVID-19 case in Brazil was on 25th February.\cite{brazil_so_what} 

Additional filtering of the data was applied for onset-to-ICU-admission, onset-to-hospital-admission and onset-to-death in order to eliminate bias introduced by potentially erroneous entries identified in the data for these distributions. We removed the rows where admission to the hospital or ICU or death happened on the same day as onset of symptoms, assuming that these were actually incorrectly inputted entries. The decision to test removing the first day is motivated firstly by the observation of a number of conspicuous data entry errors in the database, and secondly by anomalous spikes corresponding to same-day events observed in these distributions. An example of the anomalous spikes in the onset-to-death distribution is shown in Appendix Figure \ref{fig: onsetDeathErrors} for selected states.

Sensitivity analyses on data inclusion, regarding the removal of anomalous spikes in first-day data indicative of reporting errors (e.g.~in onset to hospital admission), and regarding the sensitivity of the dataset to time-series censoring effects, are set out in the Results Section \ref{subsec:sensitivity}.

 A summary of the data, including number and a range of samples per variable from the SIVEP-Gripe dataset is given in Table \ref{tab: VariablesSummary}. A breakdown of the number of data samples per state is provided in Appendix Table \ref{tab: numberSamples}.
 

Basic exploratory analysis to explain geographic variation observed in time-delay distributions adopts \emph{GeoSES} (\emph{Índice Socioeconômico do Contexto Geográfico para Estudos em Saúde}) \cite{geoses}, which measures Brazilian socioeconomic characteristics through an index composed of education, mobility, poverty, wealth,  deprivation, and segregation. We investigate correlations between the \emph{GeoSES} indicators and the time-delay means that we estimate at the state level.  Additionally,  we consider  correlations with the mean age of the population of the state and the percentage of people living in urban areas, data we obtained from \emph{Instituto Brasileiro de Geografia e Estatística} (\emph{IBGE}).\cite{ibge_population}


\subsection{Model fitting}
\begin{table}[]
\caption{Summary of the distribution data extracted from SIVEP-Gripe database.\cite{SIVEP} Number of samples ($N_\text{samples}$) is given for the whole country.\label{tab: VariablesSummary}}
\begin{tabular}{lccl}
        \toprule
Distribution              & $N_\text{samples}$ & Range (days) & \\
        \hline
Onset-to-death               & 59,271    & 1-114   &  \\
Hospital-admission-to-death  & 52,821    & 0-99    &  \\
ICU-stay                  & 21,709    & 0-89    & \\
Onset-to-hospital-admission  & 141,618   & 1-129   &   \\
Onset-to-hospital-discharge  & 69,478    & 0-120   & \\
Onset-to-ICU-admission       & 46,617    & 0-101   &  \\
Onset-to-diagnosis (PCR)       & 156,558   & 0-129   &  \\
Onset-to-diagnosis (non-PCR)   & 19,438    & 0-102   &   \\

    \hline
\end{tabular}
\end{table}
 Gamma, Weibull, log-normal, generalised log-normal,\cite{general_lognormal} and generalised gamma\cite{stacy1962generalization} PDFs are fitted to several epidemiological distributions, with the specific parameterisations provided in Appendix Section 
 \ref{subsec: ModelSelection}. 
 The parameters of each distribution are fitted in a joint Bayesian hierarchical model with partial pooling, using data from the 26 states and one federal district of Brazil, extracted and filtered to identify specific epidemiological distributions such as onset-to-death, ICU-stay, and so on.
 
As an example consider fitting a gamma PDF for the onset-to-death distribution. The gamma distribution for the $i^{\text{th}}$ state is given by
\begin{equation}
    \text{Gamma}(\alpha_i, \beta_i)\,,
\end{equation}
where shape and scale parameters are assumed to be positively constrained, normally distributed random variables
\begin{equation}
    \alpha_i \sim N(\alpha_\text{Brazil},\sigma_1)
\end{equation}
and
\begin{equation}
    \beta_i \sim N(\beta_\text{Brazil},\sigma_2)\,.
\end{equation}
The parameters $\alpha_\text{Brazil}$ and $\beta_\text{Brazil}$ denote the national level estimates, and
\begin{equation}
\sigma_1 \sim N^+(0,1)\,,\,
\sigma_2 \sim N^+(0,1)\,\,,
\end{equation}
where $N^+(\cdot)$ is a truncated normal distribution.
In this case, parameters $\alpha_\text{Brazil}$ and $\beta_\text{Brazil}$ are estimated by fitting a gamma PDF to the fully pooled data, that is including the observations for all states.
Prior probabilities for the national level parameters for each of the considered PDFs are chosen to be $N^+(0,1)$, except for the generalised gamma distribution where we used: $\mu_\text{Brazil} \sim N^+(2,0.5)$, $\sigma_\text{Brazil} \sim N^+(0.5,0.5)$ and $s_\text{Brazil} \sim N^+ (1.5,0.5)$.

Posterior samples of the parameters in the model are generated using Hamiltonian Monte Carlo (HMC) with Stan.\cite{carpenter2017stan, hoffman2014no} For each fit we use $4$ chains and $2,000$ iterations, with half of the iterations dedicated to warm-up.

The preference for one fitted model over another is characterised in terms of the Bayesian support, with the model evidence calculated to see how well a given model fits the data, and comparison between two models using Bayes Factors. The details of how to estimate the model evidence and calculate the Bayes Factors for each pair of models are given in Appendix Section \ref{subsec: ModelSelection}.

\begin{figure*}[!ht]
\begin{centering}
\includegraphics[scale=0.6]{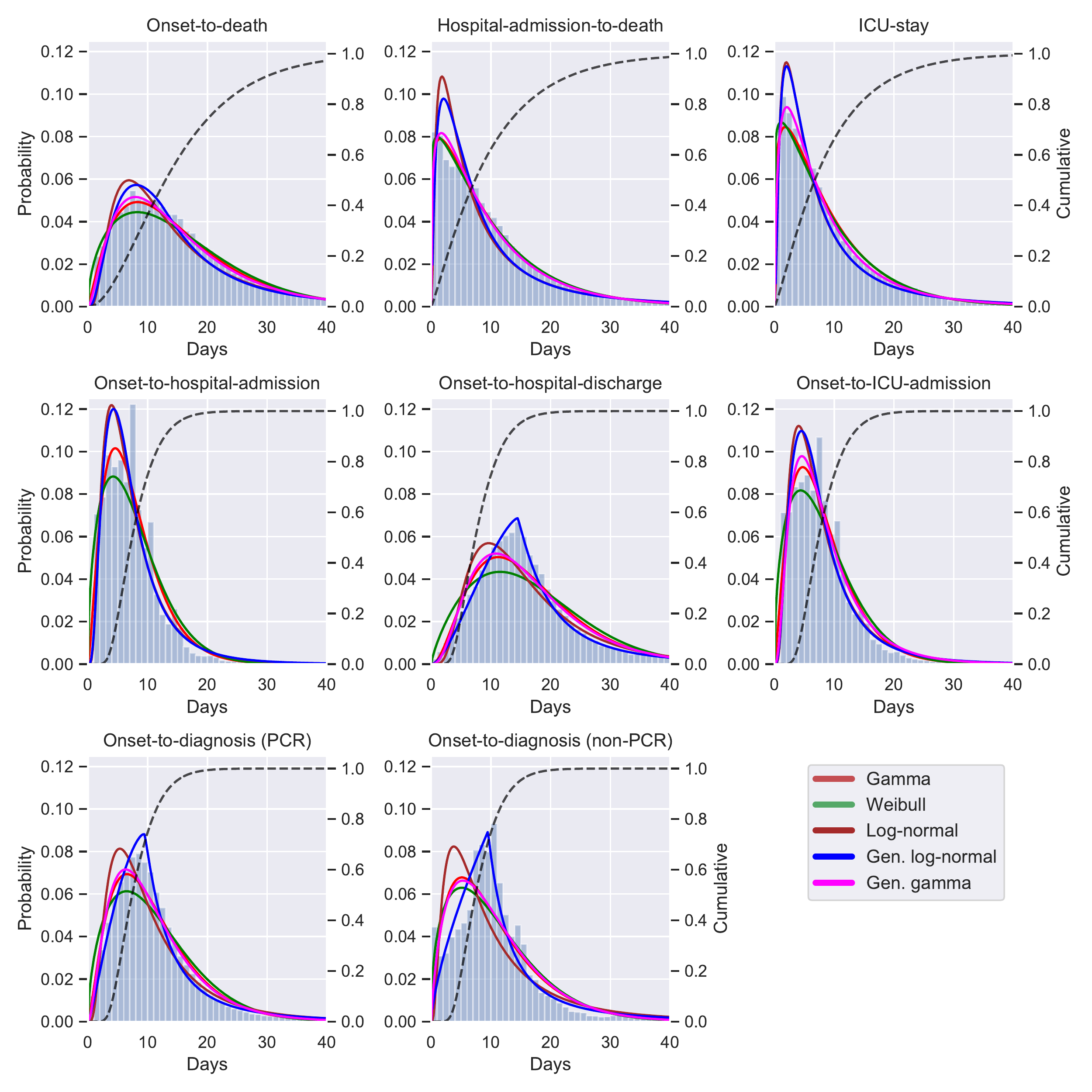}
\par\end{centering}
\caption{Histograms for onset-to-death, hospital-admission-to-death, ICU-stay, onset-to-hospital-admission, onset-to-hospital-discharge, onset-to-ICU-admission
onset-to-diagnosis (PCR) and onset-to-diagnosis (non-PCR) distributions show data for Brazil extracted from the SIVEP-Gripe database.\cite{SIVEP} For each distribution, solid lines are for fitted PDFs and the dashed line shows the cumulative distribution function of the best-fitting PDF. The left hand side y-axis gives the probability value for the PDFs and the right hand side y-axis shows the value for the cumulative distribution function. All values on the x-axes are given in days. State-level fits are shown in Figure \ref{fig: BoxMeansAllStates} and Appendix Figures \ref{fig: ridgeMeans1} and \ref{fig: ridgeMeans2}. \label{fig: AllFitsNational}}
\end{figure*}

\section{Results\label{sec:Results}}

\subsection{Brazil epidemiological distributions\label{hospital_dist}}
Five trial PDFs --  gamma, Weibull, log-normal, generalised log-normal and generalised gamma -- were fitted to the epidemiological data shown in Figure \ref{fig: AllFitsNational}. 
 
All of the models' fits were tested by using the Bayes Factors based on the Laplace approximation and corrected using thermodynamic integration,\cite{Mellan_Hawryluk_refTI, meng1996simulating, gelman1998simulating} as described in Appendix Section \ref{subsec: ModelSelection}. The thermodynamic integration contribution was negligible suggesting the posterior distributions are satisfactorily approximated as multivariate normal. The conclusions on the preferred PDF were not sensitive to the choice of prior distributions, that is the preferred model was still the favoured one even when more informative prior distributions were applied for all PDFs. The Bayes Factors used for model selection are shown in Appendix Table \ref{tab: Laplace-bayes-factors}.

The gamma PDF provided the best fit to the onset-to-death, hospital-admission-to-death and ICU-stay data. For the remaining distributions -- onset-to-diagnosis (non-PCR), onset-to-diagnosis (PCR), onset-to-hospital-discharge, onset-to-hospital-admission and onset-to-ICU-admission -- the generalised log-normal distribution was the preferred model. The list of preferred PDFs for each distribution, together with the estimated mean, variance and PDFs' parameter values for the national fits are given in Table \ref{tab: summaryResultsNational}. The 95\% credible intervals (CrI) for parameters of each of the preferred PDFs was less than 0.1 wide, therefore in Table \ref{tab: summaryResultsNational} we show only point estimates.

Additionally, in Figure \ref{fig: AllFitsNational}, in each instance the cumulative probability distribution is given for the best model fit, revealing that out of patients for whom COVID-19 is terminal, almost $70$\% die within $20$ days of symptom onset. Out of patients who die in the hospital, almost $60$\% die within the first 10 days since admission.

The estimated mean number of days for each distribution for Brazil is compared in Table \ref{tab: AllMeans} with values found in the literature for China, US and France. The majority of the data obtained through searching the literature pertained to the early stages of the epidemic in China, and no data was found for low- and middle-income countries. The mean onset-to-death time of $15.2$ (95\% CrI $15.1-15.3$) days, from a best-fitting gamma PDF, is shorter than the $17.8$ (95\% CrI $16.9–19.2$) days estimate from Verity et al.,\cite{verity_estimates_2020} and $20.2$ (95\% CrI $15.1 - 29.5$) days estimate ($14.5$ days without truncation) from Linton et al.\cite{linton2020incubation} In both cases, estimates were based on a small sample size from the beginning of the epidemic in China.
The mean number of days for hospital-admission-to-death of $10.8$ (95\% CrI $10.7 - 10.9$) for Brazil matches closely the 10 days estimated by Salje et al.\cite{china-onset-diagnosis}

\begin{table*}[!htbp]
\caption{For each COVID-19 distribution the preferred PDF with the largest Bayesian support is listed, along with the estimated mean, variance and other parameter of the PDF. 95\% credible intervals are given in brackets for mean and variance. The parameters $p_1$, $p_2$ and $p_3$ for the preferred PDFs gamma and generalised log-normal (GLN) are given in the form $Gamma(x|p_1,p_2) = Gamma(\alpha, \beta)$ and $GLN(x|p_1,p_2,p_3) = GLN(\mu, \sigma, s)$, with the formulae of the PDFs given in Appendix Section \ref{sec:Appendix}. The credible intervals for parameters $p_1$, $p_2$ and $p_3$ are less than 0.1 wide, so only the point estimates are shown. $\dagger$ The variance diverges for the onset-to-diagnosis (non-PCR) PDF.\label{tab: summaryResultsNational}}
    \centering
    \begin{tabular}{@{}ccccccc@{}}
        \toprule
        Distribution & Preferred PDF & Mean (days) & Variance (days$^2$) & $p_1$  & $p_2$  & $p_3$  \\
            \hline
        Onset-to-death & Gamma & 15.2 (15.1, 15.3) & 105.3 (103.7, 106.9) & 2.2 & 0.1 & - \\
        Hospital-admission-to-death & Gamma & 10.0 (9.9, 10.0) & 84.8 (83.2, 86.4) & 1.2 & 0.1 & - \\
        ICU-stay & Gamma & 9.0 (8.9, 9.1) & 64.9 (63.1, 66.8) & 1.2 & 0.1 & - \\
        Onset-to-hospital-admission & Gen. log-normal & 7.8 (7.7, 7.8) & 35.7 (35.0, 36.5) & 1.8 & 0.6 & 1.8 \\
        Onset-to-hospital-discharge & Gen. log-normal & 17.6 (17.6, 17.7) & 248.7 (233.7, 265.6) & 2.7 & 0.3 & 1.2 \\
        Onset-to-ICU-admission & Gen. log-normal & 8.5 (8.4, 8.5) & 48.0 (46.1, 50.0) & 1.9 & 0.6 & 1.8 \\
        Onset-to-diagnosis (PCR) & Gen. log-normal & 12.5 (12.5, 12.6) & 252.3 (236.4, 269.6) & 2.3 & 0.3 & 1.2 \\
        Onset-to-diagnosis (non-PCR) & Gen. log-normal & 14.5 (14.3, 14.7) & $\dagger$ & 2.3 & 0.3 & 1.0 \\
            \hline
    \end{tabular}
\end{table*}

\begin{figure*}[!htbp]
\begin{centering}
\includegraphics[scale=0.55]{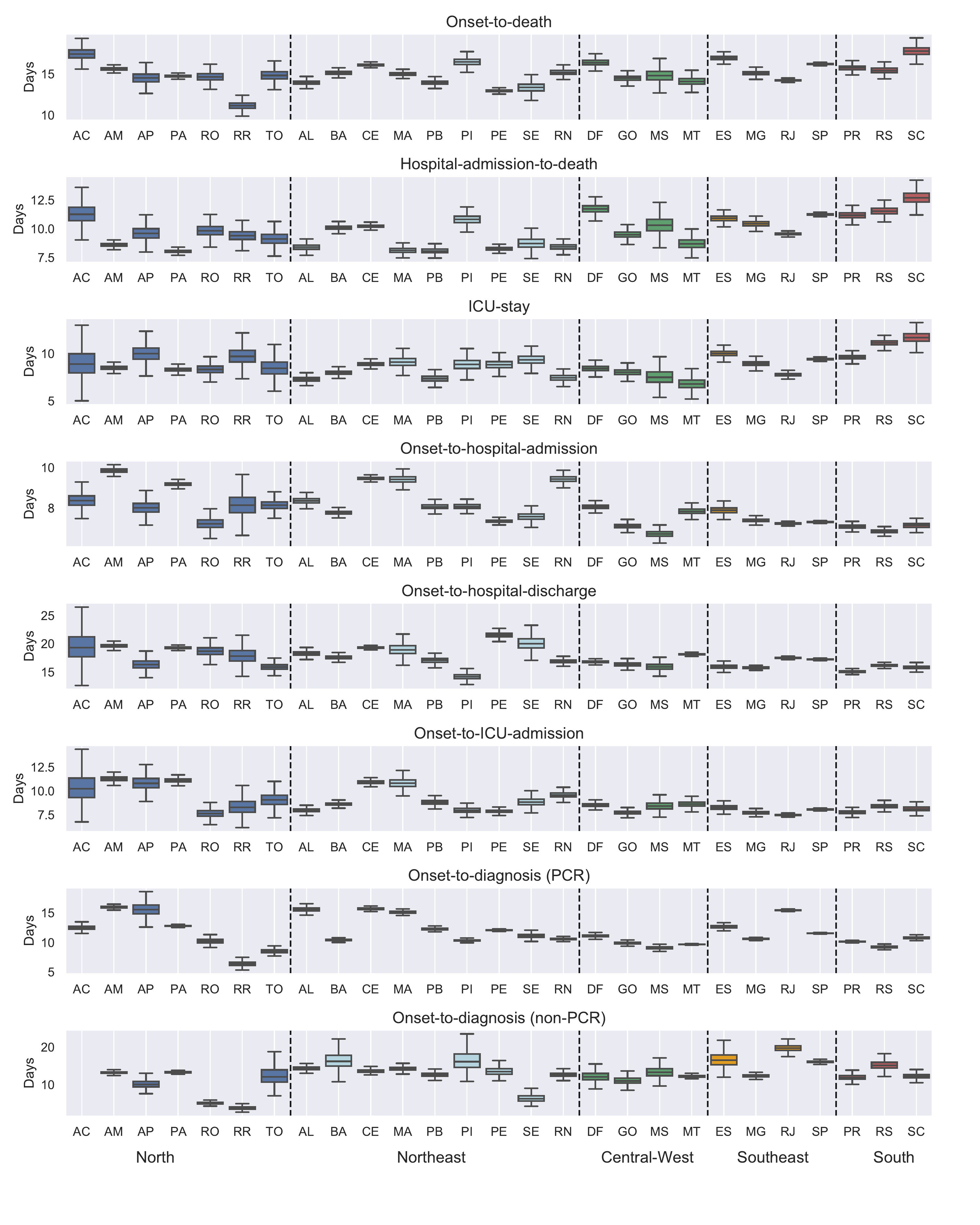}
\par\end{centering}
 \caption{
Estimates of the mean time in days for onset-to-death, hospital-admission-to-death and each of the other distributions fitted in the joint model of Brazil. Estimates are grouped by the five regions of Brazil, North (blue), Northeast (light-blue), Central-West (green), Southeast (orange), South (red), and are shown for Acre (AC), Amazonas (AM), Amapá (AP), Pará (PA), Rondônia (RO), Roraima (RR), Tocantins (TO), Alagoas (AL), Bahia (BA), Ceará (CE), Maranhão (MA), Paraíba (PB), Piauí (PI), Pernambuco (PE), Sergipe (SE), Rio Grande do Norte (RN), Distrito Federal (DF), Goiás (GO), Mato Grosso do Sul (MS), Mato Grosso (MT), Espírito Santo (ES), Minas Gerais (MG), Rio de Janeiro (RJ), São Paulo (SP), Paraná (PR),  Rio Grande do Sul (RS), Santa Catarina (SC). For state Acre, the onset-to-diagnosis (non-PCR) mean diverged due to the small number of samples (n=1). The full posterior distribution for each mean estimate is given in Appendix Figures \ref{fig: ridgeMeans1} and \ref{fig: ridgeMeans2}. \label{fig: BoxMeansAllStates}}
\end{figure*}

\subsection{Subnational Brazilian epidemiological distributions\label{hospital_dist}}
The onset-to-death distribution, and other time-delay distributions such as onset-to-diagnosis, length of ICU stay, onset-to-hospital-admission, onset-to-hospital-discharge, onset-to-ICU-admission, and hospital-admission-to-death, have been fitted in a joint model across the 26 states and one federal district of Brazil using partial pooling. The mean number of days, plotted in Figure \ref{fig: BoxMeansAllStates}, shows substantial subnational variability -- e.g. the mean onset-to-hospital-admission for Amazonas state was estimated to be 9.9 days (95\% CrI 9.7-10.1), whereas for Mato Grosso do Sul the estimate was 6.7 (95\% CrI 6.4-7.1) days and Rio de Janeiro - 7.2 days (95\% CrI 7.1-7.3). Amazonas state had the longest average time from onset-to-hospital- and ICU-admission. The state with the shortest average onset-to-death time was Acre. Santa Catarina state on the other hand had a longest average onset-to-death and hospital-admission-to-death time, as well as longest average ICU-stay. For a visualisation of the uncertainty in our mean estimates for each state, see the posterior density plots in Appendix Figures \ref{fig: ridgeMeans1} and \ref{fig:  ridgeMeans2}. Additional national and state-level results for the onset-to-death gamma PDF, including the posterior plots for mean and variance, are shown in Figure \ref{fig: onsetToDeathGamma} in the Appendix.

We also observe discrepancies between the five geographical regions of Brazil, for example states belonging to the southern part of the country (Paraná, Rio Grande do Sul and Santa Catarina) had a longer average ICU-stay and hospital-admission-to-death time as compared to the states in the North region. Full results, including detailed estimates of mean, variance, and estimates for each of the distributions' parameters for Brazil and Brazilian states can be accessed at \url{https://github.com/mrc-ide/Brazil_COVID19_distributions/blob/master/results/results_full_table.csv}.

\begin{table*}[!htb]
\caption{Epidemiological distributions for COVID-19 have been fitted for Brazil, and sources worldwide have been obtained from the literature. PDF means for Brazil have been obtained using Markov Chain Monte Carlo (MCMC) sampling, using the PDF with the maximum Bayesian support for each data distribution (see Appendix Table \ref{tab: Laplace-bayes-factors}). All values are given in days, and 95\% CrI are given in brackets unless stated otherwise. $^*$ adjusted for censoring, $\dag$ PCR confirmed, $\ddag$ non-PCR confirmed, $^a$ median (interquartile range), $^b$ mean (standard deviation). \label{tab: AllMeans}}
    \centering
    \begin{tabular}{@{}lcccc@{}}
        \toprule
        Distribution    & Brazil           & China            & France       & US    \\
        \hline
        Onset-to-death     & \begin{tabular}{c}15.2 (15.1, 15.3) \\ 16.0$^*$ (15.9, 16.1) \end{tabular} & \begin{tabular}{c}17.8 (16.9, 19.2)\cite{verity_estimates_2020} \\ 18.8$^*$ (15.7, 49.7)\cite{verity_estimates_2020} \\ 14.5 (12.5, 17.0)\cite{linton2020incubation} \\ 20.2$^*$ (15.1, 29.5)\cite{linton2020incubation} \end{tabular}  &  & 13.59$^b$ (7.85)\cite{us-onset-death} \\
        \cline{2-5}
        Hospital-admission-to-death & \begin{tabular}{c} 10.0 (9.9, 10.0) \\ 10.8$^*$ (10.7, 10.9) \end{tabular} & \begin{tabular}{c}5.0$^a$ (3.0, 9.3) \cite{china-onset-admission} \\ 8.9 (7.3-10.4)\cite{linton2020incubation} \\ 13.0$^*$ (8.7-20.9)\cite{linton2020incubation} \end{tabular}    &     10.0\cite{france} &   \\
        \cline{2-5}
        ICU-stay        & \begin{tabular}{c}9.0 (8.9, 9.1) \\ 10.1$^*$ (9.9, 10.2)\end{tabular}    & 8.0$^a$ (4.0, 12.0)\cite{china-icu-stay}   & 17.6 (17.0, 18.2)\cite{france} &\\    
        \cline{2-5}
        Onset-to-hospital-admission & 7.8 (7.7, 7.8)    & 10.0$^a$ (7.0-12.0) \cite{china-onset-admission}  &         &     \\
        Onset-to-hospital-discharge & 17.6 (17.6, 17.7) & 22.0$^a$ (18.0, 25.0) \cite{china-icu-stay} &        &      \\
        Onset-to-ICU-admission       & 8.5 (8.4, 8.5)    & 9.5$^a$ (7.0, 12.5) \cite{china-onset-icu}   &       &       \\
        \cline{2-5}
        Onset-to-diagnosis & \begin{tabular}{c} 12.5\dag (12.5, 12.6) \\ 14.5\ddag (14.3, 14.7) \end{tabular}       & 5.5 (5.4, 5.7)\cite{china-onset-diagnosis}    &         &     \\
    \hline      
    \end{tabular}
\end{table*}


\subsection{Sensitivity analyses\label{subsec:sensitivity}}

In order to remove the potential bias towards shorter outcomes from left- and right-censoring, we tested the scenario in which the data to fit the models was truncated. For example, based on a 95\% quartile of 35 days for the hospital-admission-to-death distribution, entries with the starting date (hospital admission) after 2nd June 2020 and those with an end-date (death) before 1st April were truncated, and the models were refitted. With censored parts of the data removed, the mean time from start to outcome increased for every distribution, e.g. for hospital-admission-to-death it increased from 10.0 (95\% CrI 9.9-10.0) to 10.8 (95 \% CrI 10.7-10.9), and for onset-to-death it changed from 15.2 days (95\% CrI 15.1-15.3) to 16.0 days (95\% CrI 15.9-16.1). The effect truncation on censored data is given in Appendix Figure \ref{fig: sensitivityPlot}.

To test the impact of keeping or removing entries identified as potentially resulting from erroneous data transcription (see the Methods Section \ref{sec:Method}), we fitted the PDFs to some of the distributions on a national level with and without those entries.
For onset-to-hospital-admission, onset-to-ICU and onset-to-death we find that generalised gamma PDF was preferred when the first day of the distribution was included, and gamma (for onset-to-death) and generalised log-normal PDFs if the first day was removed. For hospital-admission-to-death, a gamma distribution fitted most accurately when the first day was included, and Weibull when it was excluded. 
The effect of removing the first day results in means shifting to the right by approximately $1$ day for both onset-to-hospital- and ICU-admission, and by $0.5$ days for hospital-admission-to-death (see Appendix Figure \ref{fig: sensitivityPlot}).

Sensitivity analysis regarding the model selection approach is detailed in Appendix Section \ref{subsec: ModelSelection}.

\section{Discussion\label{sec:Discussion}}

We fitted multiple probability density functions to a number of epidemiological datasets, such as onset-to-death or onset-to-diagnosis, from the Brazilian SIVEP Gripe database,\cite{SIVEP} using Bayesian hierarchical models. 
Our findings provide the first reliable estimates of the various epidemiological distributions for the COVID-19 epidemic in Brazil and highlight a need to consider a wider set of specific parametric distributions. Instead of relying on the ubiquitous gamma or log-normal distributions, we show that often these PDFs do not best capture the behaviour of the data. For instance, the generalised log-normal  is preferable for several of the epidemiological distributions in Table \ref{tab: summaryResultsNational}. These results can inform modelling of the epidemic in Brazil \cite{mellan2020report}, and other low- and middle-income countries,\cite{walker-science-lmic} but we expect they also have some relevance more generally.

\begin{figure}[!hb]
\begin{centering}
\includegraphics[scale=0.6]{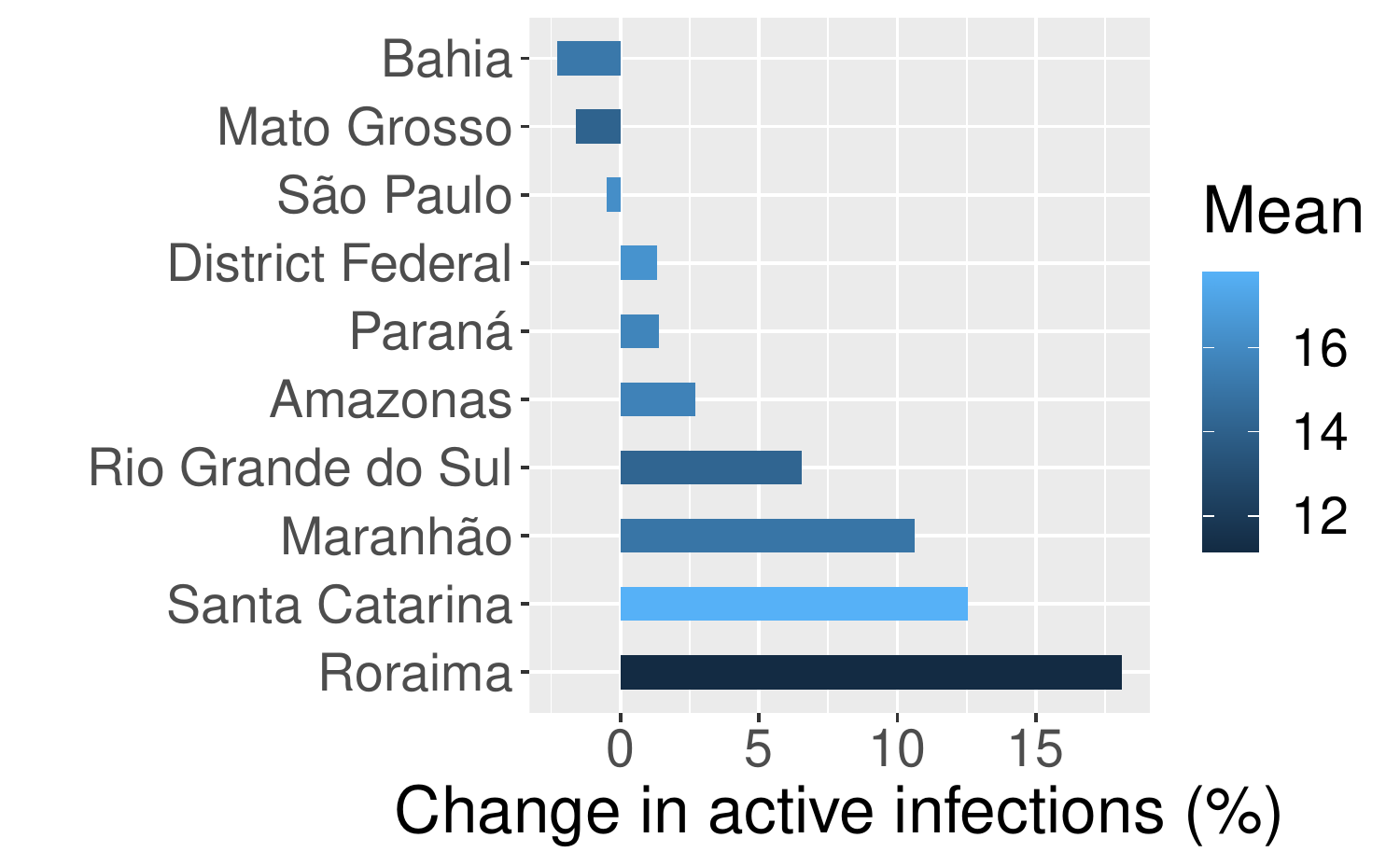}
\par\end{centering}
\caption{This figure shows the percentage change in active infections, estimated on the 23rd-Jun-2020, that results from using state-specific onset-to-death distributions (see Appendix Table \ref{tab: StatesOnsetDeath}) compared to a single national-level one. The effect for each state is coloured according to the mean of the state's onset-to-death gamma distribution, given in days. The mean onset-to-death for Brazil is $15.2$ days.
\label{fig: ai_bar_chart}}
\end{figure}

In terms of modelling the epidemic in Brazil, the variation observed at subnational level -- see Figure \ref{fig: BoxMeansAllStates} -- can be shown to be important to accurately estimating disease progression. Making use of the state-level custom-fitted onset-to-death distributions reported here, we have estimated the number of active infections on 23rd June 2020 across ten states spanning the five regions of Brazil, using a Bayesian hierarchical renewal-type model.\cite{flaxman2020nature,mellan2020report,mishra2020derivation} The relative change in the number of active infections from modelling the cases using heterogeneous state-specific onset-to-death distributions, compared to using a single common Brazil one is shown in Figure \ref{fig: ai_bar_chart} to be quite substantial. The relative changes observed, up to 18\% more active infections, suggest assumptions of onset-to-death homogeneity are unreliable and closer attention needs to be paid when fitting models of SARS-CoV-2 transmission dynamics in large countries.

On the origin of the geographic variation displayed in Figure \ref{fig: BoxMeansAllStates} for the average distribution times across states, there are multiple potential factors that could generate the observed variability and in this work we present an elementary exploratory analysis. We examine the correlation between socioeconomic factors, such as education, poverty, income, etc., using a number of socioeconomic state-level indicators obtained from Barrozo et al.(2020) \cite{geoses} and additional datasets containing the mean age per state and percentage of people living in the urban areas (urbanicity).\cite{ibge_population} The Pearson correlation coefficients, shown in the Appendix Table \ref{tab: socEcoCorr}, suggest that segregation, poverty and deprivation elements were most strongly correlated with the analysed onset-time datasets. E.g. poverty was strongly negatively correlated with hospital-admission-to-death (-0.68), whereas income and segregation had a high positive correlation coefficient for the same distribution (+0.60, +0.62 respectively). The strongest correlation was observed for hospital-admission-to-death and deprivation indicator, which measures the access to sanitation, electricity and other material and non-material goods.\cite{geoses} Interestingly, the indicators measuring economical situation were more correlated with average hospitalisation times than mean age per state, which suggests that although the low- and middle-income countries typically have younger populations, their healthcare systems are more likely to struggle in response to the COVID-19 epidemic. More detailed analysis is necessary to fully appreciate the impact of the economic components on the COVID-19 epidemic response. 


In the work presented we acknowledge numerous limitations. The database from which distributions have been extracted, though extensive, contains transcription errors, and the degree to which these bias our estimates is largely unknown. Secondly, the PDFs fitted are based on observational hospital data, and therefore should be cautiously interpreted for other settings. Thirdly, though we have fitted PDFs at subnational as well as national level, this partition is largely arbitrary and further work is required to understand the likely substantial effect of age, sex, ethnic variation,\cite{sivep-ethnic} co-morbidities, and other factors.

\section{Conclusions\label{sec:Conclusions}}

We provide the first estimates of common epidemiological distributions for the COVID-19 epidemic in Brazil, based on the SIVEP-Gripe hospitalisation data.\cite{SIVEP} Extensive heterogeneity in the distributions between different states is reported. Quantifying the time-delay for COVID-19 onset and hospitalisation data provides useful input parameters for many COVID-19 epidemiological models, especially those modelling the healthcare response in low- and middle-income countries.

\section{Acknowledgements}
We thank Microsoft for providing Azure credits which were used to run the analysis.

\section{Funding}
This work was supported by Centre funding from the UK Medical Research Council under a concordat with the UK Department for International Development, the NIHR Health Protection Research Unit in Modelling Methodology and Community Jameel. This research was also partly funded by the Imperial College COVID-19 Research Fund. IH was supported by Imperial College London MRC Centre.

\section{Code and Data availability}
Python, R and Stan code used to analyse the data and fit the distribution is available at \url{https://github.com/mrc-ide/Brazil_COVID19_distributions}, along with estimated parameters for each state and PDFs considered at \url{https://github.com/mrc-ide/Brazil_COVID19_distributions/blob/master/results/results_full_table.csv}. The SIVEP-Gripe database,\cite{SIVEP} is available to download from Brazil Ministry of Health website \url{https://opendatasus.saude.gov.br/dataset/bd-srag-2020}.

\clearpage
\section{References}
\bibliography{ref.bib}

\clearpage

\section{Appendix \label{sec:Appendix}}
\begin{table*}[!h]
\caption{Probability density functions with analytical formulae for mean and variance. $y$ denotes the data, $\Gamma(\cdot)$ is a gamma function. GG -- generalised gamma, GLN -- generalised log-normal. 
\label{tab: PDF}}
\begin{tabular}{lll}
\toprule
PDF & Mean & Variance \\
\hline
$\text{Gamma}(y|\alpha, \beta) = \frac{\beta^\alpha}{\Gamma(\alpha)}y^{\alpha-1}\exp(-\beta y)$ & $\frac{\alpha}{\beta}$ & $\frac{\alpha}{\beta^2}$ \\
$\text{Weibull}(y|\alpha, \sigma) = \frac{\alpha}{\sigma} \left(\frac{y}{\sigma}\right)^{\alpha-1} \exp\left(-\left(\frac{y}{\sigma}\right)^\alpha\right)$ & $\sigma\Gamma\left(1+\frac{1}{\alpha}\right)$ & $\sigma^{2}\left(\Gamma\left(1+\frac{2}{\alpha}\right)-\Gamma^{2}\left(1+\frac{1}{\alpha}\right)\right)$ \\
$\text{Log-normal}(y|\mu, \sigma) = \frac{1}{\sqrt{2\pi}\sigma}\frac{1}{y}\exp\left(-\frac{1}{2}\left(\frac{\log y-\mu}{\sigma}\right)^2\right)$ & $\text{exp}\left(\mu+\frac{\sigma^{2}}{2}\right)$ & $\left(\text{exp}\left(\sigma^{2}\right)-1\right)\exp\left(2\mu+\sigma^{2}\right)$ \\
$\text{GG}(y|a, d, p) = \frac{1}{\Gamma\left(\frac{d}{p}\right)}\left(\frac{p}{a}\right)^d x^{d-1} \exp\left(-\left(\frac{y}{a}\right)^p\right)$ & $a\frac{\Gamma\left(\left(d+1\right)/p\right)}{\Gamma\left(d/p\right)}$ & $a^{2}\left[\frac{\Gamma\left(\left(d+2\right)/p\right)}{\Gamma\left(d/p\right)}-\left(\frac{\Gamma\left(\left(d+2\right)/p\right)}{\Gamma\left(d/p\right)}\right)^{2}\right]$ \\
$\text{GLN}(y|\mu, \sigma, s) = \frac{1}{y} \frac{s}{2^{\frac{s+1}{s}}\sigma\Gamma\left(\frac{1}{s}\right)} \exp\left(-\frac{1}{2}|\frac{\log y-\mu}{\sigma}|^s\right)$ & \begin{tabular}{l}\text{exp\ensuremath{\left(\mu\right)\left[\ensuremath{1+\frac{1}{2\Gamma\left(1/s\right) } \cdot S}\right]},} \\ \ensuremath{S = \sum_{j=1}^{\infty}\sigma^{j}\left(1+\left(-1\right)^{j}\right)2^{j/s}\frac{\Gamma\left(\frac{j+1}{2}\right)}{\Gamma\left(j+1\right)}} \end{tabular}
 & \begin{tabular}{l}\text{exp\ensuremath{\left(2\mu\right)\left[\ensuremath{1+\frac{1}{2\Gamma\left(1/s\right) } \cdot S}\right]}}-\text{\ensuremath{\left[ \text{Mean}\right]^{2}},} \\ \ensuremath{S = \sum_{j=1}^{\infty}2\sigma^{j}\left(1+\left(-1\right)^{j}\right)2^{j/s}\frac{\Gamma\left(\frac{j+1}{2}\right)}{\Gamma\left(j+1\right)}} \end{tabular} \\
\hline
\end{tabular}
\end{table*}

\subsection{Model selection\label{subsec: ModelSelection}}
To characterise which model (gamma, log-normal, etc.) best fits the data, the Bayesian model evidence $z = z(y|M_i)$  is evaluated.
Here and throughout this section $y$ denotes the data and $M_i$ denotes the $i^\text{th}$ model from the analysed model set. As determining the model evidence requires calculating an integral over the model parameters ($\theta$) which is generally intractable, we approximate it with $z_0 = z_0(y|M_i)$, which is based on a second-order Laplace approximation,\cite{laplace} $q_0 = q_{0}(\theta|M_i,y)$, to the true un-normalised posterior density $q = q(\theta|M_i,y)$. The second-order approximated density is estimated as:
\begin{equation}
q_{0} = q(\hat{\theta})\,\exp \left(-\frac{1}{2}\, (\mathrm{\bm{\theta}} - \hat{\bm{\theta}}) \, \Sigma^{-1} \, (\bm{\theta} - \bm{\hat{\theta}})^T\right)\,.
\label{eq: LaplaceReference}
\end{equation}
Here $q(\hat{\theta})$ denotes the value of the un-normalised posterior evaluated using the mean estimates of the model's parameters $\hat{\theta}$, and $\Sigma$ the covariance matrix built from Markov Chain Monte Carlo (MCMC) samples of the posterior distribution. From this expression, a second-order approximation to the model evidence, $z_0$, is given by $z_0 =  q(\hat{\theta})\sqrt{\text{det}(2\pi\Sigma^{-1})}$, where $\text{det}(\cdot)$ denotes the determinant of the matrix.

For each model pair, Bayes factors were computed from the marginal likelihoods. Considering two models $M_i$ and $M_j$, the Bayes Factor (BF) is
\begin{equation}
B_{ij} = \frac{z(y|M_i)}{z(y|M_j)}\,,
\label{eq: BayesFactor}
\end{equation}
where $z(y|M_i)$ is the evidence of model $M_i$ given $y$.
If $B_{ij} > 1$, the evidence is in favour of model $M_i$. Here, for readability we will report the Bayes Factors as $2\,\text{log}(B_{ij})$ following Kass and Raftery notation.\cite{bayes_factors}

The sensitivity of our model evidence is tested with respect to the choice of hyperprior distribution, and secondly with respect to the use of the approximate second-order density $q_0$. In the latter instance this is done by performing thermodynamic integration\cite{Mellan_Hawryluk_refTI, meng1996simulating, gelman1998simulating} between $q_0$ and the true density $q$ in order to obtain an asymptotically exact estimate of the marginal model evidence,
\begin{talign}
z = \textrm{\ensuremath{z_{0}\,}exp}\left(\int_{0}^{1}\mathbb{E}_{\theta\sim q(\theta;\,\lambda)}\left[\textrm{log}\,q_{}-\textrm{log}\,q_{0}\right]d\lambda\,\right).
\label{eq: thermodynamicIntegral}
\end{talign}
 The right hand term corrects the $z_0$ approximation to the exact Bayesian evidence by a path integral evaluated with respect to a sampling distribution that interpolates between the two densities as $q(\theta;\,\lambda) = q^{(1-\lambda)}q_0^\lambda$ in terms of the auxiliary coordinate $\lambda$.

\begin{table*}[h]
\caption{ Bayes Factors (BFs) for the analysed distributions and models. For each distribution (rows), the values represent BF for the best fitting model against other models. Value of 0 indicates the model that fits the data the best. Value $>$ 10 indicates a very strong evidence against given model compared to the best one. GLN - generalised log-normal, GG - generalised gamma. NA - not analysed. The BF values are reported here as $2\,\text{log}(B_{ij})$ following Kass and Raftery notation.\cite{bayes_factors} \label{tab: Laplace-bayes-factors}}
\begin{tabular}{llllll}
\toprule
                        & Gamma & Weibull & Log-normal & GLN  & GG          \\
\hline
Onset-to-death              & 0     & 2156     & 2208      & 198  & 301         \\
Admission-death      & 0     & 195     & 4349       & 3096 & 188         \\
ICU stay                & 0     & 231     & 588        & 607  & 352         \\
Onset-to-hospital-admission & 4000   & 17073     & 494      & 0   & NA           \\
Onset-to-hospital-discharge & 2819   & 8346    & 6079      & 0    & 3087         \\
Onset-to-ICU-admission      & 798   &  4359     & 142      & 0 & 1244 \\
Onset-to-diagnosis (PCR)     & 1111   & 10400     & 13882 & 0    & 1257         \\
Onset-to-diagnosis (non-PCR)  & 578   & 793     & 4340   & 0    & 461         \\

\hline
\end{tabular}
\end{table*}

\begin{table*}[h]
\caption{State-level onset-to-death estimates for gamma PDF: mean, variance, parameters values, with 95\% confidence intervals. The parameters $p_1$ and $p_2$ are given in the form $Gamma(x|p_1,p_2) = Gamma(\alpha, \beta)$. The full PDFs for other distributions are available at \url{https://github.com/mrc-ide/Brazil_COVID19_distributions/blob/master/results/results_full_table.csv}. \label{tab: StatesOnsetDeath}}
    \centering
    \begin{tabular}{@{}lllll@{}}
        \toprule
        State & Mean (days) & Variance (days$^2$) & $p_1$ & $p_2$ \\
        \hline
        AC & 17.4 (16.1, 18.8) & 119.4 (98.8, 143.6) & 2.6 (2.2, 2.9) & 0.1 (0.1, 0.2) \\
        AL & 14.0 (13.4, 14.5) & 82.5 (74.3, 91.9) & 2.4 (2.2, 2.5) & 0.2 (0.2, 0.2) \\
        AM & 15.6 (15.3, 16.0) & 95.3 (89.1, 102.1) & 2.6 (2.4, 2.7) & 0.2 (0.2, 0.2) \\
        AP & 14.5 (13.2, 16.0) & 99.1 (79.8, 122.7) & 2.1 (1.9, 2.4) & 0.1 (0.1, 0.2) \\
        BA & 15.1 (14.7, 15.6) & 116.6 (107.9, 126.1) & 2.0 (1.9, 2.1) & 0.1 (0.1, 0.1) \\
        CE & 16.1 (15.8, 16.4) & 116.4 (111.1, 122.0) & 2.2 (2.2, 2.3) & 0.1 (0.1, 0.1) \\
        DF & 16.4 (15.6, 17.2) & 105.0 (92.7, 119.0) & 2.6 (2.3, 2.8) & 0.2 (0.1, 0.2) \\
        ES & 17.0 (16.4, 17.5) & 107.8 (98.2, 118.1) & 2.7 (2.5, 2.9) & 0.2 (0.1, 0.2) \\
        GO & 14.5 (13.8, 15.2) & 87.9 (77.9, 99.1) & 2.4 (2.2, 2.6) & 0.2 (0.2, 0.2) \\
        MA & 15.0 (14.6, 15.4) & 89.4 (82.7, 96.5) & 2.5 (2.4, 2.7) & 0.2 (0.2, 0.2) \\
        MG & 15.1 (14.6, 15.7) & 95.1 (86.3, 104.7) & 2.4 (2.2, 2.6) & 0.2 (0.1, 0.2) \\
        MS & 14.8 (13.3, 16.4) & 93.9 (74.8, 116.8) & 2.4 (2.0, 2.7) & 0.2 (0.1, 0.2) \\
        MT & 14.1 (13.1, 15.1) & 80.6 (67.2, 96.4) & 2.5 (2.2, 2.8) & 0.2 (0.2, 0.2) \\
        PA & 14.7 (14.5, 15.0) & 90.2 (85.7, 94.9) & 2.4 (2.3, 2.5) & 0.2 (0.2, 0.2) \\
        PB & 14.0 (13.4, 14.5) & 78.7 (71.2, 87.3) & 2.5 (2.3, 2.7) & 0.2 (0.2, 0.2) \\
        PE & 13.0 (12.7, 13.2) & 89.7 (84.6, 95.1) & 1.9 (1.8, 1.9) & 0.1 (0.1, 0.2) \\
        PI & 16.5 (15.6, 17.4) & 114.8 (99.4, 131.7) & 2.4 (2.1, 2.6) & 0.1 (0.1, 0.2) \\
        PR & 15.7 (15.1, 16.4) & 91.9 (81.8, 102.7) & 2.7 (2.5, 2.9) & 0.2 (0.2, 0.2) \\
        RJ & 14.2 (14.0, 14.4) & 103.3 (99.5, 107.3) & 2.0 (1.9, 2.0) & 0.1 (0.1, 0.1) \\
        RN & 15.2 (14.6, 15.9) & 91.9 (81.8, 103.0) & 2.5 (2.3, 2.7) & 0.2 (0.2, 0.2) \\
        RO & 14.7 (13.6, 15.8) & 92.1 (76.4, 110.0) & 2.3 (2.1, 2.6) & 0.2 (0.1, 0.2) \\
        RR & 11.2 (10.2, 12.1) & 68.1 (55.9, 83.0) & 1.8 (1.6, 2.1) & 0.2 (0.1, 0.2) \\
        RS & 15.4 (14.7, 16.2) & 116.0 (103.0, 130.8) & 2.1 (1.9, 2.2) & 0.1 (0.1, 0.1) \\
        SC & 17.8 (16.7, 19.0) & 146.8 (125.1, 173.5) & 2.2 (1.9, 2.4) & 0.1 (0.1, 0.1) \\
        SE & 13.4 (12.2, 14.5) & 112.5 (91.4, 138.6) & 1.6 (1.4, 1.8) & 0.1 (0.1, 0.1) \\
        SP & 16.2 (16.0, 16.4) & 114.8 (111.6, 118.0) & 2.3 (2.2, 2.3) & 0.1 (0.1, 0.1) \\
        TO & 14.8 (13.5, 16.2) & 97.3 (79.1, 119.7) & 2.3 (2.0, 2.6) & 0.2 (0.1, 0.2) \\
        Brazil & 15.2 (15.1, 15.3) & 105.3 (103.7, 106.9) & 2.2 (2.2, 2.2) & 0.1 (0.1, 0.1) \\
        \hline
    \end{tabular}
\end{table*}

\begin{figure*}[]
\begin{centering}
\includegraphics[scale=0.7]{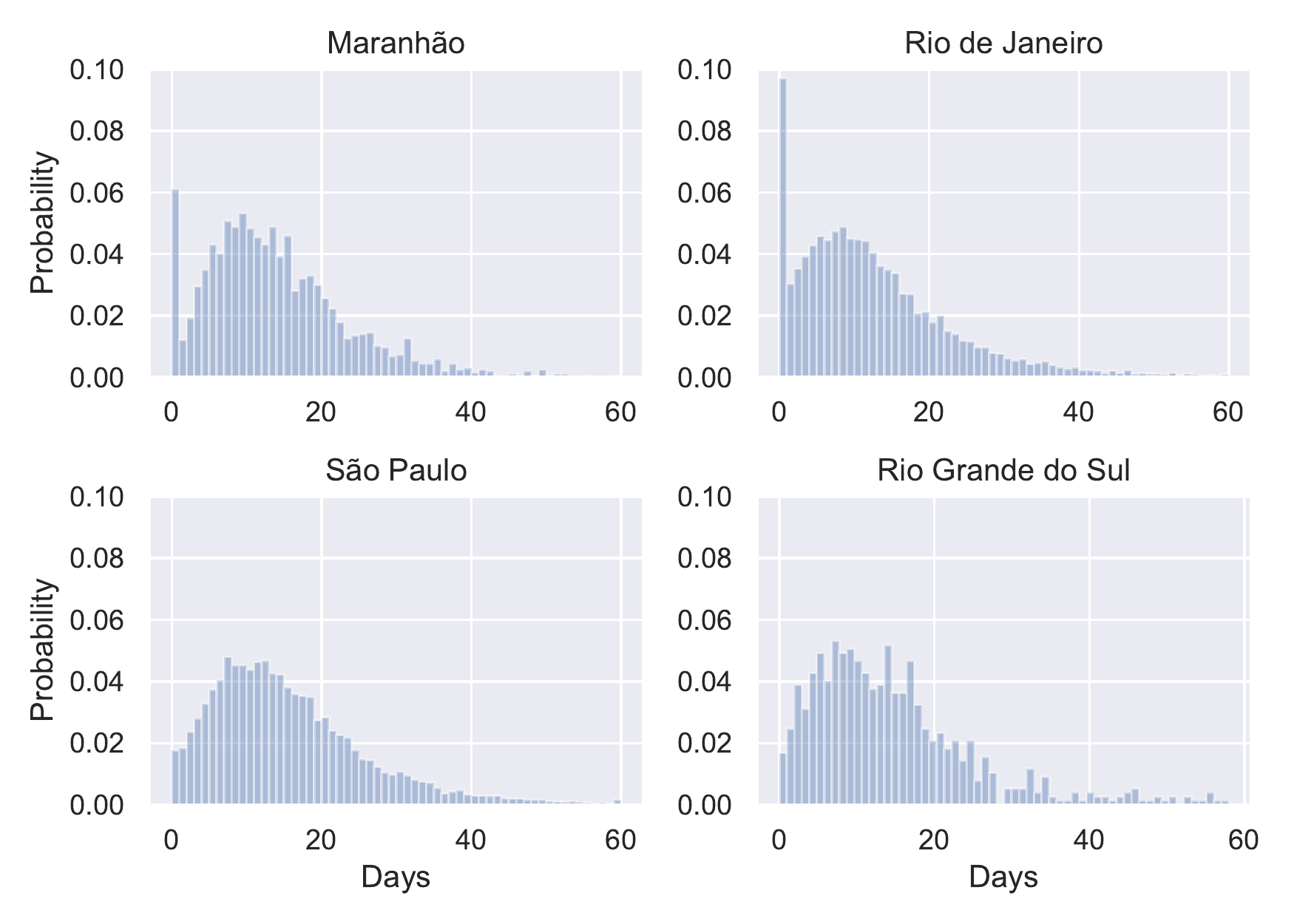}
\par\end{centering}
\caption{Distribution of onset-to-death for Maranhão, Rio de Janeiro, São Paulo and Rio Grande do Sul. Anomalous spikes for the first day can be observed for Maranhão and Rio de Janeiro, indicating they might be a reporting error. \label{fig: onsetDeathErrors}}
\end{figure*}

\begin{figure*}[]
\begin{centering}
\includegraphics[scale=0.7]{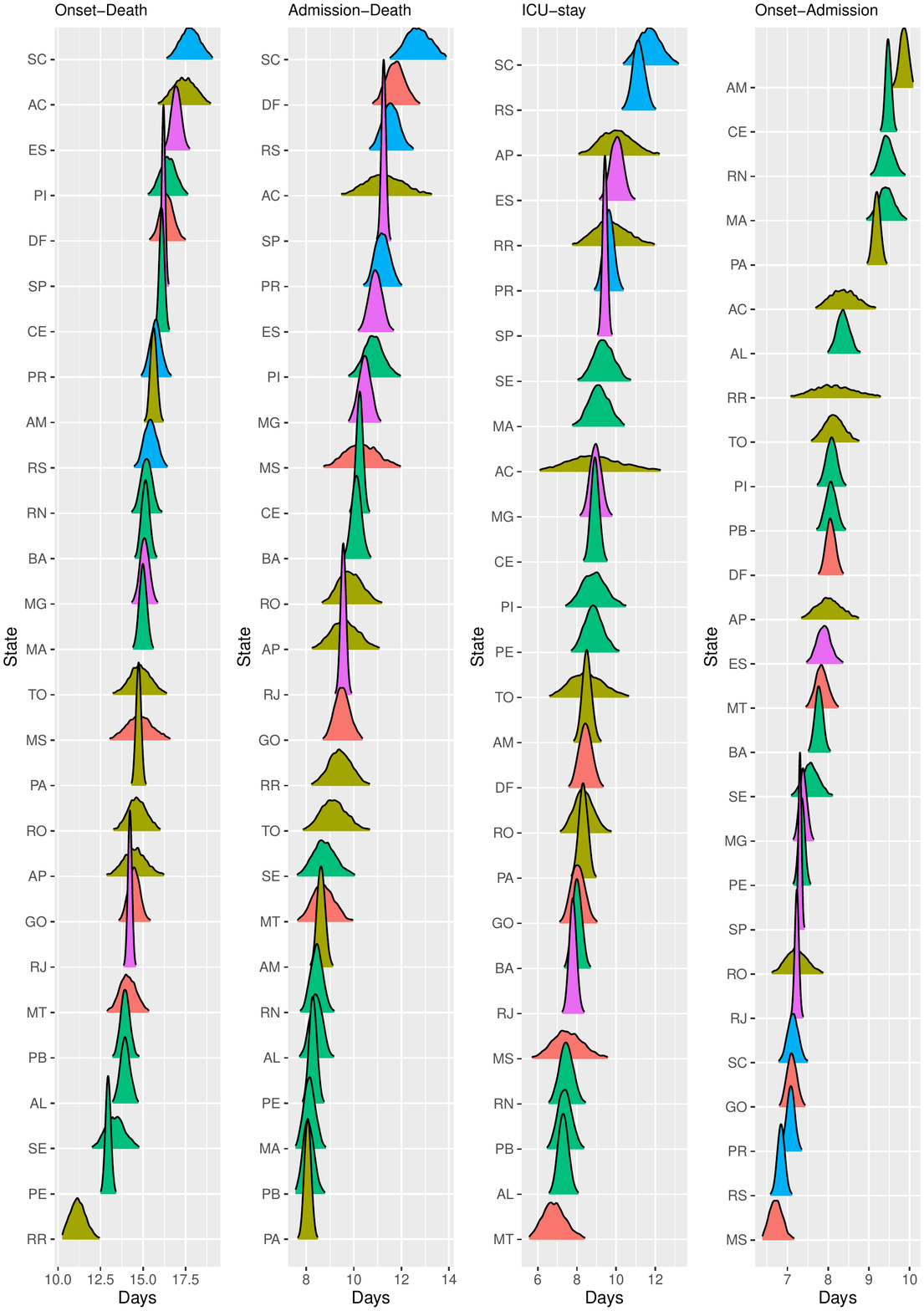}
\par\end{centering}
\caption{Posterior distribution of mean times (in days) for onset-to-death, hospital-admission-to-death, ICU stay and onset-to-hospital-admission, sorted by mean value. Plots are colour-coded by the geographical region which the state belongs to: North (yellow), Northeast (green), Central-West (orange), Southeast (purple), South (blue).   \label{fig: ridgeMeans1}}
\end{figure*}

\begin{figure*}[]
\begin{centering}
\includegraphics[scale=0.7]{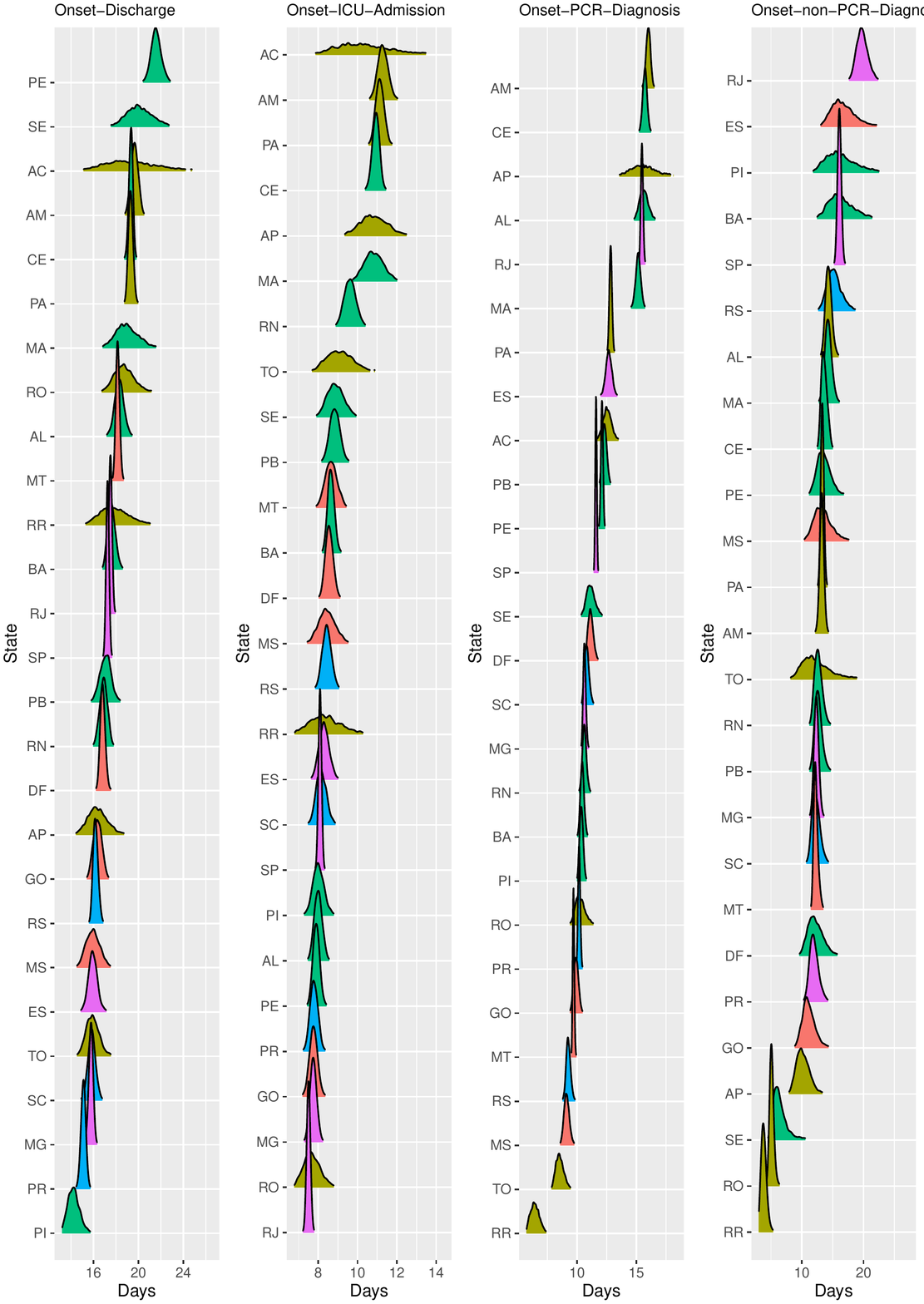}
\par\end{centering}
\caption{Posterior distribution of mean times (in days) for onset-to-hospital-discharge, onset-to-ICU-admission, onset-to-diagnosis (PCR) and onset-to diagnosis (non-PCR), sorted by mean value. Plots are colour-coded by the geographical region which the state belongs to: North (yellow), Northeast (green), Central-West (orange), Southeast (purple), South (blue).\label{fig: ridgeMeans2}}
\end{figure*}

\begin{figure*}[]

\begin{centering}
\includegraphics[scale=0.5]{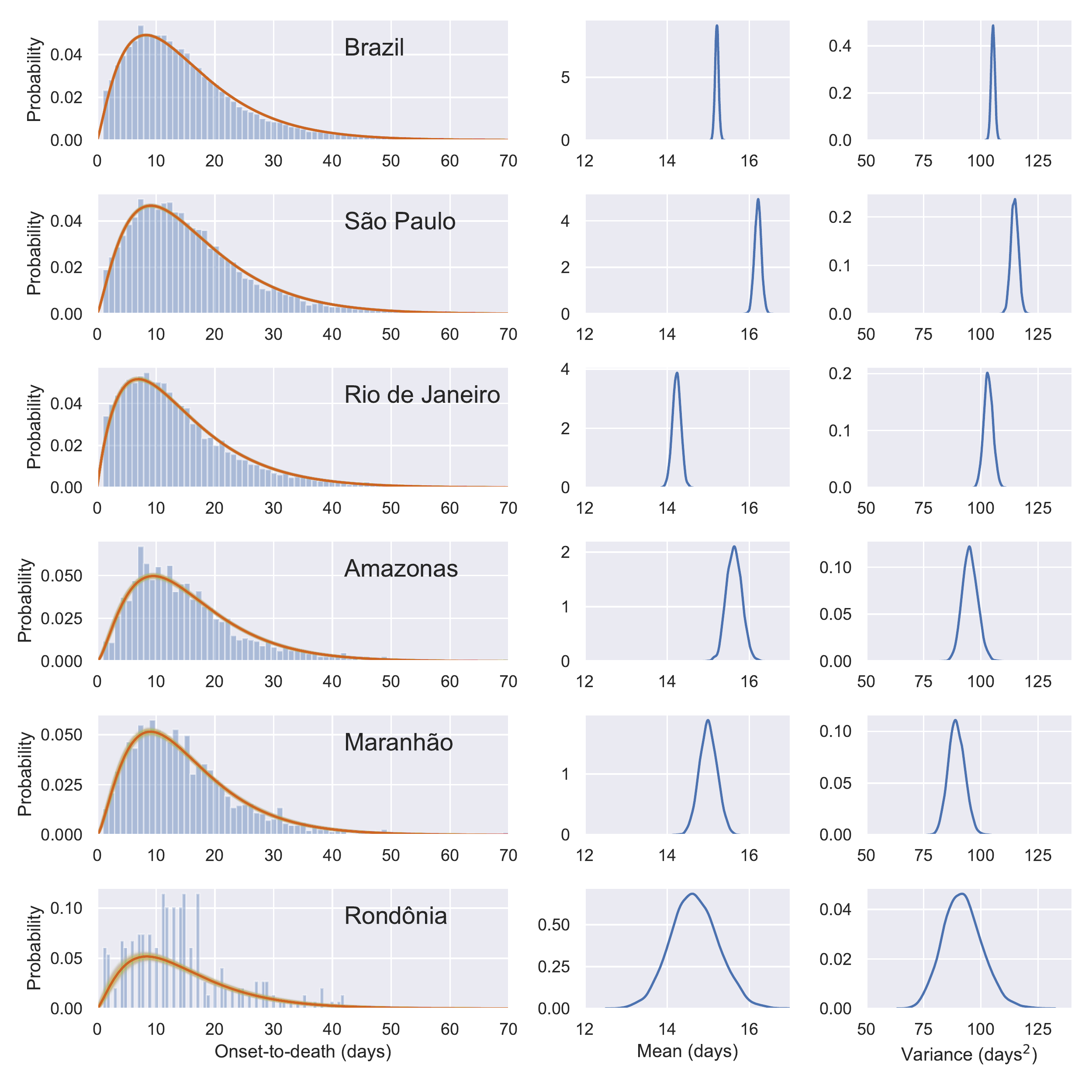}
\par\end{centering}
\caption{Gamma PDF $Gamma(\alpha, \beta)$ fitted to the onset-to-death data for Brazil and five states of Brazil. The PDFs were fitted with HMC partially pooling each state with the whole country. The red lines represent the model using the mean parameter estimates. Individual PDFs selected during MCMC sampling are shown in yellow. Posterior mean and variance distributions for each region are given in the middle and right hand side columns. \label{fig: onsetToDeathGamma}}
\end{figure*}

\begin{figure*}[h]
\begin{centering}
\includegraphics[scale=0.6]{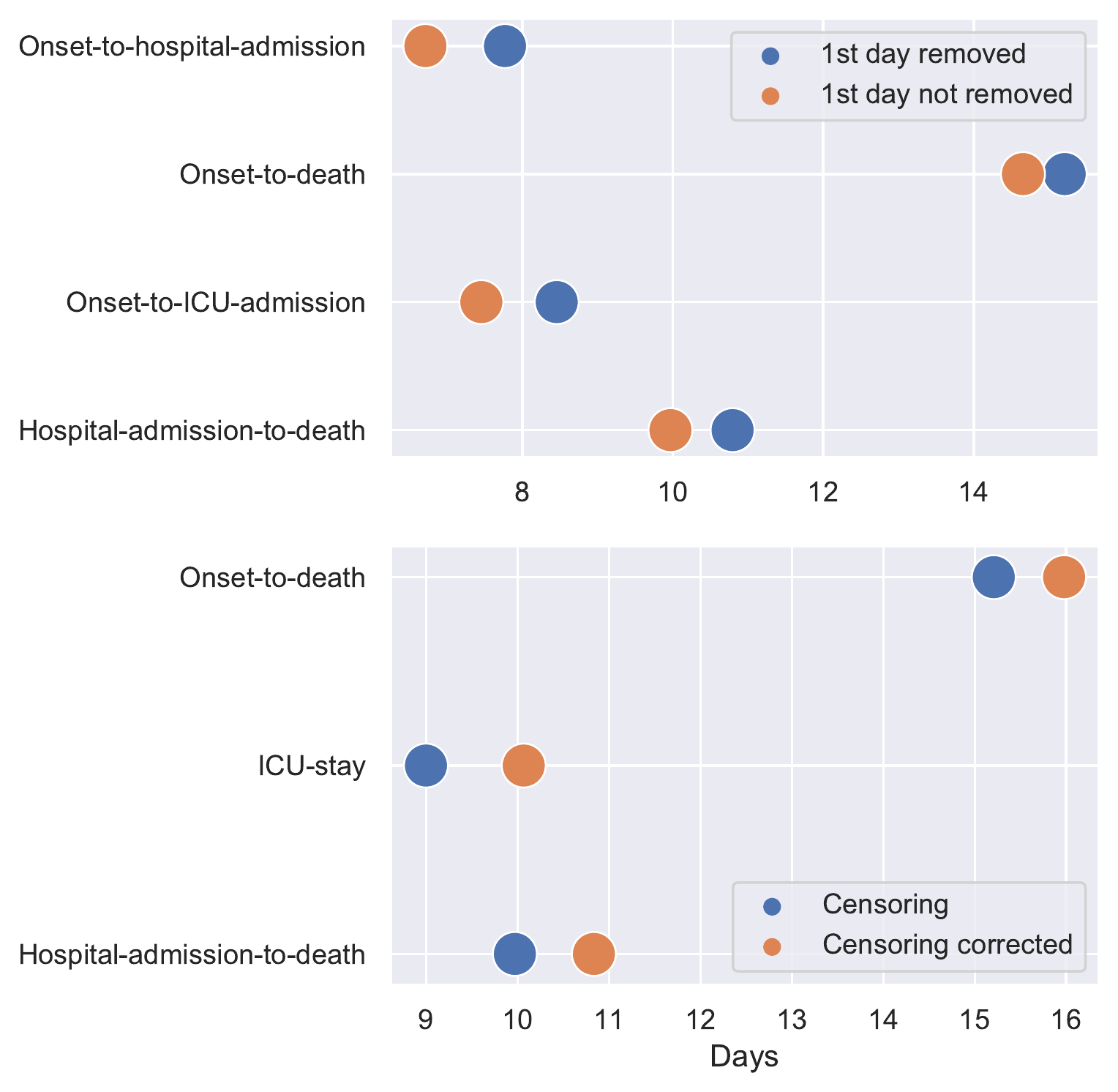}
\par\end{centering}
\caption{Estimated mean per distribution in different scenarios: excluding 1st day data points (top) and censoring correcting (bottom). The credible intervals were not shown as due to the large amount of data available they were negligible.\label{fig: sensitivityPlot}}
\end{figure*}

\begin{table*}[h] 
\centering 
  \caption{Pearson correlation coefficients for mean distribution times and socioeconomic indicators. Sample size was equal to 27 (number of states).} \label{tab: socEcoCorr} 
\begin{tabular}{cccccccc} 
\hline 
\hline
 & ICU-stay & Onset-death & Admission-death & Onset-discharge & \begin{tabular}[c]{@{}c@{}}Onset-hospital\\admission\\ \end{tabular} & \begin{tabular}[c]{@{}c@{}}Onset-ICU\\admission\\ \end{tabular} & \begin{tabular}[c]{@{}c@{}}Onset-diagnosis\\(PCR) \end{tabular} \\ 
\hline
Education &   -0.32  &   -0.25  &   -0.62  &  0.41  &  0.48  &  0.39  &  0.34  \\ 
Poverty &   -0.31  &   -0.31  &   -0.68  &  0.52  &  0.69  &  0.54  &  0.49  \\ 
Deprivation &  0.38  &  0.35  &  0.71  &   -0.49  &   -0.59  &   -0.49  &   -0.41  \\ 
Wealth &   -0.08  &  0.26  &  0.37  &   -0.24  &   -0.07  &   -0.21  &   -0.17  \\ 
Income &  0.21  &  0.28  &  0.60  &   -0.35  &   -0.40  &   -0.33  &   -0.35  \\ 
Segregation &  0.40  &  0.35  &  0.62  &   -0.43  &   -0.57  &   -0.47  &   -0.30  \\ 
Mean age &  0.13  &  0.25  &  0.43  &   -0.45  &   -0.57  &   -0.68  &   -0.25  \\ 
Urbanicity &  0.12  &  0.11  &  0.43  &  -0.34  &  -0.52  &  -0.40  &  -0.19  \\
\hline
\end{tabular} 
\end{table*}

\begin{table*}[h] \centering 
  \caption{Pearson correlation coefficients for mean distribution times. Sample size was equal to 27 (number of states).} 
  \label{tab: distrCorr} 
\begin{tabular}{ccccccc} 
\hline 
\hline 
 & Onset-death & Admission-death & Onset-discharge & \begin{tabular}[c]{@{}c@{}}Onset-hospital\\admission\\ \end{tabular} & \begin{tabular}[c]{@{}c@{}}Onset-ICU\\admission\\ \end{tabular} & \begin{tabular}[c]{@{}c@{}}Onset-diagnosis\\(PCR) \end{tabular} \\ 
\hline
Onset-death &  1  &  0.69  &   -0.35  &  0.06  &  0.24  &  0.15  \\ 
Admission-death &  0.69  &  1  &   -0.52  &   -0.48  &   -0.20  &   -0.36  \\ 
Onset-discharge &   -0.35  &   -0.52  &  1  &  0.39  &  0.43  &  0.40  \\ 
Onset-to-hospital-admission &  0.06  &   -0.48  &  0.39  &  1  &  0.72  &  0.53  \\ 
Onset-to-ICU-admission &  0.24  &   -0.20  &  0.43  &  0.72  &  1  &  0.50  \\ 
Onset-to-diagnosis (PCR) &  0.15  &   -0.36  &  0.40  &  0.53  &  0.50  &  1  \\
\hline
\end{tabular} 
\end{table*}

\begin{table*}[h]
\caption{Number of datapoints per state for each of the datasets analysed in the study. Acre (AC), Amazonas (AM), Amapá (AP), Pará (PA), Rondônia (RO), Roraima (RR), Tocantins (TO), Alagoas (AL), Bahia (BA), Ceará (CE), Maranhão (MA), Paraíba (PB), Piauí (PI), Pernambuco (PE), Sergipe (SE), Rio Grande do Norte (RN), Distrito Federal (DF), Goiás (GO), Mato Grosso do Sul (MS), Mato Grosso (MT), Espírito Santo (ES), Minas Gerais (MG), Rio de Janeiro (RJ), São Paulo (SP), Paraná (PR),  Rio Grande do Sul (RS), Santa Catarina (SC).   \label{tab: numberSamples}}
    \centering
    \begin{tabular}{@{}ccccccccc@{}}
        \toprule
          & Onset-death & Admission-Death & ICU-stay & \begin{tabular}[c]{@{}c@{}}Onset-hospital\\admission\\ \end{tabular} &
          \begin{tabular}[c]{@{}c@{}}Onset-hospital\\discharge\\ \end{tabular} &
          \begin{tabular}[c]{@{}c@{}}Onset-ICU\\admission\\ \end{tabular} &
          \begin{tabular}[c]{@{}c@{}}Onset-diagnosis\\(PCR) \end{tabular} &
          \begin{tabular}[c]{@{}c@{}}Onset-diagnosis \\(non-PCR) \end{tabular} \\
        \hline
         AC & 239 & 115 & 2 & 225 & 4 & 9 & 345 & 1 \\
        AL & 1040 & 894 & 680 & 1600 & 629 & 859 & 1344 & 416 \\
        AM & 2736 & 2403 & 1010 & 5971 & 2573 & 1323 & 4502 & 1604 \\
        AP & 181 & 175 & 68 & 299 & 136 & 80 & 183 & 153 \\
        BA & 2241 & 2013 & 982 & 4563 & 1338 & 2300 & 5266 & 352 \\
        CE & 5801 & 4905 & 1534 & 9685 & 4536 & 2768 & 8286 & 1749 \\
        DF & 662 & 655 & 499 & 2687 & 1415 & 1198 & 2864 & 311 \\
        ES & 1292 & 1023 & 589 & 1409 & 507 & 778 & 1774 & 321 \\
        GO & 698 & 637 & 375 & 1813 & 783 & 819 & 2018 & 122 \\
        MA & 1950 & 1097 & 197 & 1485 & 247 & 341 & 1562 & 821 \\
        MG & 1223 & 1176 & 603 & 4782 & 2210 & 1521 & 4910 & 604 \\
        MS & 131 & 124 & 46 & 723 & 417 & 171 & 764 & 126 \\
        MT & 286 & 248 & 83 & 1347 & 2191 & 384 & 4695 & 2175 \\
        PA & 4727 & 3934 & 1270 & 8226 & 3034 & 1993 & 6921 & 1351 \\
        PB & 1136 & 1037 & 349 & 1992 & 508 & 740 & 1584 & 644 \\
        PE & 4408 & 3284 & 311 & 6574 & 1888 & 1566 & 9745 & 190 \\
        PI & 515 & 497 & 139 & 2161 & 341 & 490 & 2314 & 240 \\
        PR & 793 & 773 & 898 & 3174 & 1952 & 1168 & 3490 & 124 \\
        RJ & 9750 & 9068 & 1490 & 18019 & 7438 & 7165 & 21159 & 1446 \\
        RN & 876 & 821 & 337 & 1878 & 664 & 693 & 1517 & 544 \\
        RO & 254 & 238 & 180 & 554 & 180 & 284 & 488 & 293 \\
        RR & 270 & 265 & 53 & 98 & 51 & 56 & 200 & 92 \\
        RS & 790 & 770 & 971 & 3565 & 2328 & 1277 & 4144 & 477 \\
        SC & 408 & 389 & 291 & 1600 & 777 & 599 & 1634 & 343 \\
        SE & 303 & 295 & 193 & 938 & 181 & 306 & 1116 & 117 \\
        SP & 16348 & 15808 & 8515 & 55735 & 32937 & 17642 & 63184 & 4769 \\
        TO & 213 & 177 & 44 & 515 & 213 & 87 & 549 & 53 \\
        \hline

    \end{tabular}
\end{table*}

\end{document}